\documentclass[]{iopjournal}

\usepackage{graphicx}%
\usepackage{multirow}%
\usepackage{amsmath,amssymb,amsfonts}%
\usepackage{amsthm}%
\usepackage{mathrsfs}%
\usepackage[title]{appendix}%
\usepackage{xcolor}%
\usepackage{textcomp}%
\usepackage{manyfoot}%
\usepackage{booktabs}%
\usepackage{algorithm}%
\usepackage{algorithmicx}%
\usepackage{algpseudocode}%
\usepackage{listings}%

\usepackage{setspace}

\usepackage{isotope}
\usepackage{graphicx,tabularx,booktabs,colortbl}
\usepackage{pgfplotstable}
\usepackage{subcaption}
\usepackage[load-configurations=abbreviations, separate-uncertainty=true, multi-part-units=single]{siunitx}
\usepackage[T1]{fontenc}

\usepackage[style=vancouver,natbib=true,sorting=none]{biblatex}
\newcommand{\eq}{\; = \;}

\newcommand{\scm}{\isotope[43/44]{Sc}}
\newcommand{\scp}{\isotope[44]{Sc}}

\newcommand{\crmax}{CRC_{max}}
\newcommand{\crmean}{CRC_{mean}}
\newcommand{\crpeak}{CRC_{peak}}
\newcommand{\pg}{p$\gamma$}
\newcommand{\cov}{\mathrm{COV} }

\begin{document}
\articletype{Paper}

\title{Quantitative Imaging of Scandium Isotopes with a Long Axial Field-of-View PET/CT}

\author{Lorenzo Mercolli$^{1,2}$\orcid{0000-0002-3170-7311}, Mohammadreza Teimoorisichani$^3$, Alenka E.\ Gaschen$^{1,2}$, Pascal V.\ Grundler$^4$, Anzhelika N.\ Moiseeva$^4$, Saverio Braccini$^2$, Nicholas P.\ van der Meulen$^{4,5}$}

\affil{$^1$Department of Nuclear Medicine, Inselspital, Bern University Hospital, University of Bern, Bern, Switzerland }

\affil{$^2$Albert Einstein Center for Fundamental Physics (AEC), Laboratory for High Energy Physics (LHEP), University of Bern, Bern, Switzerland}

\affil{$^3$Siemens Medical Solutions USA, Inc., Knoxville TN, USA}

\affil{$^4$PSI Center for Life Sciences, Villigen PSI, Switzerland}

\affil{$^5$PSI Center for Nuclear Engineering and Sciences, Villigen PSI, Switzerland}

\email{lorenzo.mercolli@insel.ch}

\begin{abstract}

\emph{Objective.} \isotope[43]{Sc} and \isotope[44]{Sc} are positron-emitting isotopes of scandium that have attracted growing interest for theranostic applications in conjunction with the therapeutic isotope \isotope[47]{Sc}. Both isotopes are non-pure positron emitters, and the associated prompt-gamma emission together with the high positron energy introduce complexities in achieving precise quantitative imaging. 

\emph{Approach.} We evaluated the quantitative imaging performance of \scp\ and a \scm\ mixture on a long axial field-of-view (LAFOV) PET/CT using a NEMA IEC body phantom. 
We assessed contrast recovery coefficients, image noise and bias for different numbers of iterations and for absolute versus relative scatter correction, and implemented an isotope-specific point-spread-function (PSF) correction based on FLUKA-simulated positron ranges.

\emph{Main results.}
The quantitative imaging performance of both isotopes is comparable to \isotope[18]{F}, and the vendor's prompt-gamma correction effectively suppresses the additional random coincidences and lung error caused by prompt-gamma emission. The isotope-specific PSF correction further improves contrast recovery. While the bias obtained for \scp\ meets the EARL acceptance criterion for non-standard radionuclides, the bias for the \scm\ mixture exceeds it, which we attribute to the mixture-specific cross-calibration of the dose calibrators rather than to the PET imaging.

\emph{Significance.}
Our results support the use of LAFOV PET/CT for quantitative imaging of scandium isotopes and provide practical reconstruction recommendations for clinical protocols.
\end{abstract}

%
\vspace{2pc}

\section{Introduction}

The success story of theranostic applications of \isotope[68]{Ga}/\isotope[18]{F}- and \isotope[177]{Lu}-labeled prostate-specific membrane antigen (PSMA) ligands and DOTA-conjugated peptides \cite{Gandhi2025,Behera2026} has fostered significant efforts to extend the theranostic concept to novel radionuclide pairs that offer a close chemical match between the diagnostic and therapeutic radionuclides. Scandium isotopes have emerged as ideal theranostic radionuclides with a chemically more coherent alternative to \isotope[18]{F}/\isotope[68]{Ga}- and \isotope[177]{Lu}- labeled tracers \cite{Muller2014,Muller2018,Mikolajczak2019,vandermeulen2021,Malik2026}. 
In addition, the production of scandium isotopes can build upon the existing infrastructure of medical cyclotrons \cite{vandermeulen2020,braccini2020,Dellepiane2022,Meier2025} and there are even generator systems being developed for \scp\ \cite{Schmidt2023,Malik2026}. In the context of diagnostics and treatment planning with PSMA, it is highly desirable to have positron emission tomography (PET) radionuclides with a longer half-life than the standard \isotope[18]{F} and \isotope[68]{Ga} in order to keep good count statistics while imaging multiple or late time points of small lesions \cite{Alberts2021b}.

Both PET radioisotopes of scandium, i.e.\ \isotope[43]{Sc} and \isotope[44]{Sc}, are non-pure positron emitters \cite{Conti2016}. This means that the decay radiation contains additional prompt photons. While such prompt photons (\pg) might be useful for special applications \cite{Mercolli2025,Huang2025,Das2026}, they can lead to a higher random coincidence rate and, thereby, deteriorate the PET image \cite{Conti2016,Rosar2021,Preylowski2013,DaubeWitherspoon2021}. Moreover, the positrons from scandium decay have a relatively large energy and the \pg\ can lead to a higher radiation dose burden for the patient \cite{Gomes2024} compared to \isotope[18]{F}. Exploiting the aforementioned favorable properties of scandium therefore depends on the development and validation of reliable correction methods for the \pg\ contribution.


The isotopes \isotope[43]{Sc}  and \isotope[44]{Sc} have already been imaged with clinical PET/CT systems \cite{lima2020,lima2021,vandermeulen2021}, demonstrating their potential as positron emitters for diagnostic nuclear medicine. Although, first-in-human studies were reported in the literature \cite{sing2015,Eppard2017,Zhang2019} some time ago, scandium radioisotopes are not yet available in clinical routine. 
Building on these earlier works, the aim of the present study is two-fold. Firstly, we aim to extend the evaluation of scandium isotopes to long axial field-of-view (LAFOV) PET/CT \cite{Prenosil2022a,Prenosil2022b,Spencer2021,Zhang2024,DaubeWitherspoon2021} and to short/low-activity acquisitions. Secondly, we aim to investigate the imaging properties of \isotope[43]{Sc} in more detail.

\section{Methods}

Enriched calcium targets were irradiated at the IP2 irradiation station of the Paul Scherrer Institute (PSI, Villigen-PSI, Switzerland) and chemically processed to produce pure \scp\ as well as a mixture of \isotope[43]{Sc} and \scp\ \cite{grundler2020,vandermeulen2020}. Despite the enrichment of \isotope[44]{Ca} in the target, it is not feasible to produce isotopically pure \isotope[43]{Sc} via the [44]{Ca}(p,2n) nuclear reaction \cite{Carzaniga2017}. About \SI{240}{\mega\becquerel} of \scm\ and \SI{334}{\mega\becquerel} of \scp\ were sent from PSI to the Department of Nuclear Medicine at Inselspital (Bern, Switzerland) where a NEMA IEC body phantom (Data Spectrum Corporation, Durham, NC, USA) was filled according to the EARL standard 2 protocol \cite{Boellaard2010,Kaalep2018,Sunderland2025}. The data were acquired with a Biograph Vision Quadra (Siemens Medical Solutions USA Inc, Knoxville, TN, USA) LAFOV PET/CT scanner. All image reconstructions were performed offline with the prototype software e7-tools (Siemens Medical Solutions USA Inc, Knoxville, TN, USA). 

The dose calibrator at Inselspital (VDC-405/VIK-202, Comecer, Italy) was simply cross-calibrated with the dose calibrator from PSI \cite{Juget2023}. No calibration factor from the vendor nor from the Swiss metrology institute was provided. For the case of \scm\ this is particularly challenging as the definition of \SI{}{\becquerel} from the activity measurement is somewhat ambiguous when compared to the \SI{}{\becquerel\per\ml} that are measured in the PET/CT scan. We therefore applied the following procedure. At PSI a \SI{0.999}{\mega\becquerel} aliquot, as measured in PSI's dose calibrator, of the \scm\ solution was measured in an HPGe spectrometer. At reference time, the HPGe measurement identified \SI{0.310}{\mega\becquerel} \isotope[44]{Sc}, \SI{0.010}{\mega\becquerel} \isotope[44m]{Sc} and \SI{0.864}{\mega\becquerel} \isotope[43]{Sc}, which exceeds the total activity measured by the dose calibrator by $18.5\%$. In order to determine the syringe activity, i.e.\ the activity in the stock solution for the phantom filling, a scaling factor between PSI's and Insel's dose calibrators was computed and the (decay-corrected) isotope inventory from the HPGe measurement was rescaled accordingly (the half-life of \isotope[43]{Sc} is \SI{3.891\pm0.012}{\hour} \cite{NuDat2021} and \SI{4.042\pm0.007}{\hour} for \isotope[44]{Sc} \cite{Duran2022} ).
After removing the \isotope[44m]{Sc} activity from the inventory (there is hardly any positron emission), the sum of the \isotope[43]{Sc} and \isotope[44]{Sc} activities at scan time can be compared to the activity concentration in the PET images. The mixture-specific positron fraction was taken into account. The two phantoms were prepared with a sphere-to-background ratio of about $10:1$. Tab.~\ref{tab:activities} summarizes the activity concentrations at scan time. At scan time, the \scm\ mixture contained $72.2\%$ \isotope[43]{Sc} and $27.8\%$ \isotope[44]{Sc}, respectively.

\begin{table}[]
    \centering
    \caption{Details of the phantom preparation at scan time. For the \scm\ mixture we list the two isotopes separately. }\label{tab:activities}   
    {\footnotesize
    \begin{tabular}{lccc} 
    \toprule 
    Phantom/isotope& $A_s$ [\SI{}{\kilo\becquerel\per\ml}] & $A_{bkg}$ [\SI{}{\kilo\becquerel\per\ml}] &  Positron fraction \\ 
    \midrule 
    \rowcolor{lightgray} \phantom{Mixture} \isotope[43]{Sc} & $30.18$ & $3.03$  & $0.881$ \\  
    \rowcolor{lightgray} \phantom{Mixture} \isotope[44]{Sc} & $11.61$ & $1.17 $ & $0.943$ \\  
    \rowcolor{lightgray} \multirow{-3}*{Mixture} \scm & $41.79$ & $4.20 $ &  $0.898$ \\ 
    %
    Pure \scp & $40.45$ & $4.05$ & $0.943$ \\
    \bottomrule 
    \end{tabular}
    }
\end{table}

The scans were performed using the \isotope[43]{Sc} and \isotope[44]{Sc} settings of Quadra. The preset half-lives and positron fractions on the Quadra are \SI{3.891}{\hour} and $0.881$ for \isotope[43]{Sc} and \SI{3.97}{\hour} and $0.9427$ for \scp, respectively. We adapted the decay factors in order to correct for the most recent value of \SI{4.042\pm0.007}{\hour} half-life for \scp\ reported in Ref.~\cite{Duran2022}. While the short scan time makes the difference in decay factor negligible, the cross-calibration of the dose calibrators is significantly affected. Tab.~\ref{tab:activities} also reports the positron fraction of the \scm\ mixture at scan time. Although the difference to the pure \isotope[43]{Sc} positron fraction is marginal, we corrected the positron fraction in the images.  

Reconstructions of the NEMA phantom data were performed with the Siemens' proprietary ordered-subset expectation-maximization (OSEM) algorithm, including time-of-flight (TOF), attenuation and scatter correction, as well as point-spread-function (PSF) recovery. The list-mode data were restricted to a \SI{5}{\min} acquisition. The matrix size was $440 \times 440$ and we did not apply a post-reconstruction filter (all-pass), despite being part of the clinical protocols using Quadra \cite{Prenosil2022a,Prenosil2022b}. The random sinogram was estimated through a delayed time window. We compared different numbers of iterations and scatter-correction models (absolute vs. relative \cite{Bal2021}); five subsets were used throughout since this is the clinical standard for Quadra. It is important to note that for absolute scatter correction, no \pg\ correction is computed. In the case of relative scatter, a \pg\ sinogram is estimated by fitting a linear combination of a scaled uniform background sinogram and simulated TOF single scatter sinograms to the tail region of the emission sinogram \cite{Watson2007,Conti2016,Bal2021}.

\begin{figure}
    \centering
    \includegraphics[width=0.95\linewidth]{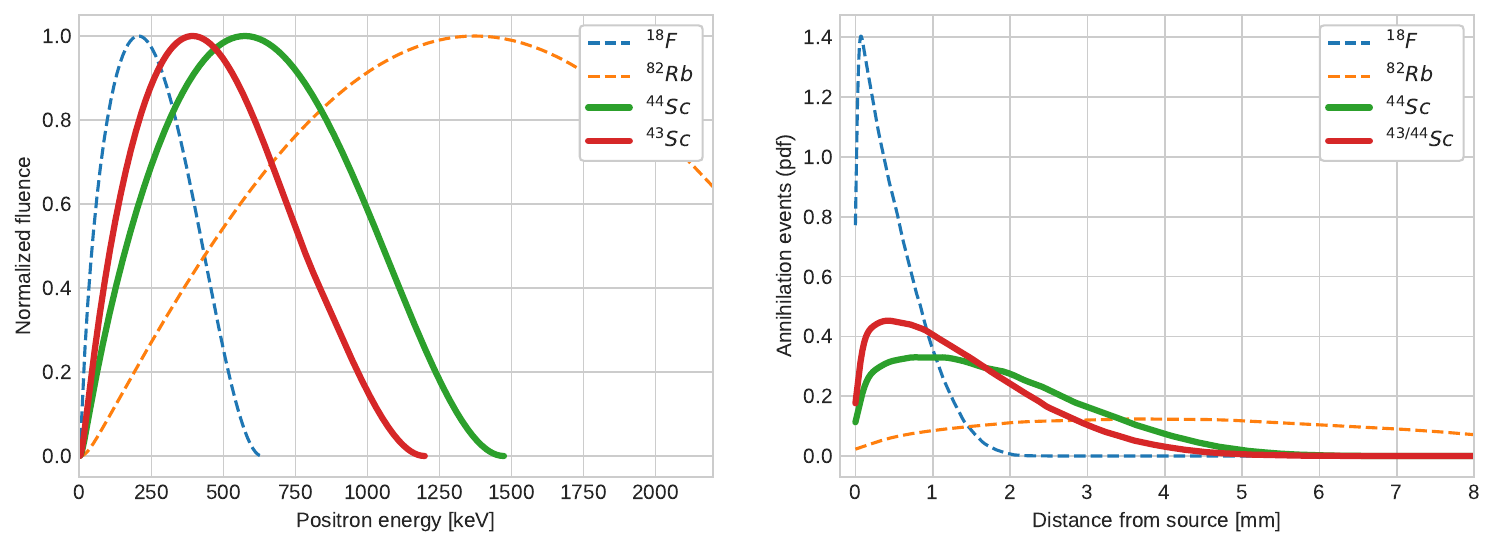}
    \caption{Positron energy spectrum (\emph{left}) and positron range in water (\emph{right}) of scandium isotopes in comparison with \isotope[18]{F} and \isotope[82]{Rb}.}
    \label{fig:positron_range}
\end{figure}

As shown in the left panel of Fig.~\ref{fig:positron_range}, the positron energy of \isotope[43]{Sc} and \scp\ is higher than that of \isotope[18]{F}, though still lower than that of \isotope[82]{Rb}. In order to correct for the increased positron range in water (see right panel of Fig.~\ref{fig:positron_range}), we computed the positron range using the FLUKA particle transport toolkit \cite{Battistoni2015,Ahdida2022,Donadon2024}. The annihilation locations were scored using a modified MGDRAW routine and the resulting kernel densities were applied in image space using a Richardson-Lucy deconvolution (same methodology as in Ref.~\cite{Mercolli2026b}). We refrained from implementing a custom positron range correction in the OSEM projectors \cite{Gavriilidis2022,Kertesz2022,Lassen2025}. However, by deconvolving only the isotope-specific part of the PSF, we preserved the geometric/scanner-specific PSF correction \cite{Panin2006}. In the following, we will refer to the images with an isotope-specific PSF as `corrected PSF' images. 

We evaluated the image properties of scandium radioisotopes according to the updated EARL standard 2 protocol \cite{Boellaard2010,Kaalep2018,Sunderland2025} for \isotope[18]{F}. In particular, we followed Ref.~\cite{Sunderland2025} and considered contrast recovery coefficients (CRC) rather than the previously-used standardized uptake value (SUV) recovery coefficients (RC). On the PET image, the maximum pixel values $C_{max}^{(i)}$ in each sphere $s_i$ $ i=1,\ldots 6$ of the NEMA phantom were identified and a background-corrected $50\%$ isocontour volume-of-interest (VOI) drawn therefrom to compute the mean activity concentration $C_{mean}^{(i)}$ in each sphere. The peak activity concentration $C_{peak}$ was computed by placing a spherical VOI with $\emptyset\SI{12}{\mm}$ diameter such that it yields the highest mean activity concentration. The contrast recovery coefficients for each sphere $i$ are therefore defined as 

\begin{equation}
    CRC^{(i)}_\mu \; \doteq \;  \left( 1-\frac{C_{\mu}^{(i)}}{C_{bkg}} \right) \cdot \left( 1-\frac{A_s^{(i)}}{A_{bkg}} \right)^{-1}  \qquad  \mbox{with} \qquad \mu \eq \{ max, \, mean, \, peak \} \;,  
\end{equation}
where $A_s$ and $A_{bkg}$ are the activity concentrations in the spheres and the background compartment of the phantom reported in Tab.~\ref{tab:activities}. The background activity concentration $C_{bkg}$ is determined through the mean activity $\bar{C}_j$ in three quadratic regions of interest (ROIs) in three distinct slices in the phantom's background compartment, i.e.\ $C_{bkg} = 1/9 \sum_{j=1}^9 \bar{C}_j$.  

The bias $B = 1-C_{bkg}/A_{bkg}$ measures the quantitative accuracy \cite{Sunderland2025}. While a $|B|\leq 5\%$ should be achieved for \isotope[18]{F} and \isotope[68]{Ga}, a deviation of up to $10\%$ is accepted for other radionuclides \cite{Sunderland2025}. The coefficient of variation (COV), i.e.\ the inverse of the signal-to-noise ratio, is defined through the standard deviation $\sigma_j$ in each background ROI.

\begin{equation}
    \mathrm{COV} \; \doteq \; \frac{1}{9} \sum_{j=1}^9 \mathrm{COV}_j \eq \frac{1}{9} \sum_{i=1}^9 \frac{\sigma_{j}}{\bar{C}_{j}} \cdot 100 \;. 
\end{equation}
$\mathrm{COV}\leq 15\%$ is compliant with the EARL limit for image noise. The residual lung error is defined as the mean activity concentration in a cylindrical VOI (\SI{120}{\mm} length, \SI{30}{\mm} diameter) placed at the center of the phantom's lung insert divided by $C_{bkg}$. Also, for the lung error $15\%$ is considered acceptable.


\section{Results}

\begin{figure}
    \centering
    \includegraphics[width=0.75\linewidth]{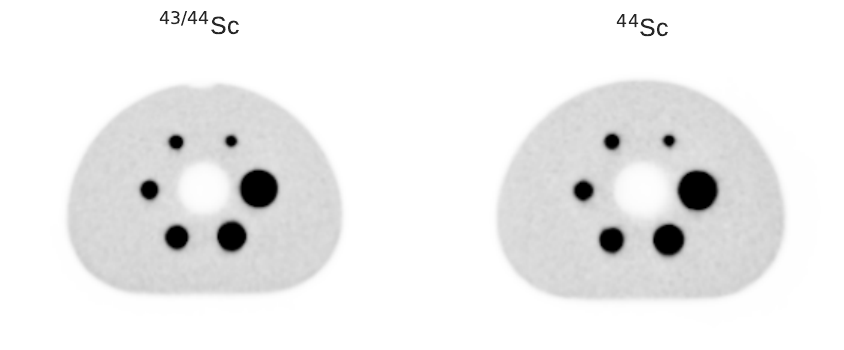}
    \caption{Example images of the NEMA phantom for \scm\ (\emph{left}) and \scp\ (\emph{right}). }
    \label{fig:phantom_slice}
\end{figure}

For illustrative purposes, Fig.~\ref{fig:phantom_slice} shows the two central slices of the \scm\ and \scp\ NEMA phantom images. These images were reconstructed with three iterations, relative scatter correction, standard PSF correction and without post-reconstruction filter.  

The bias for the two phantom scans is $B_{\scm} =-17.1\%$ and $B_{\scp} =9.6\%$ for \scp, respectively.

\begin{figure}
    \centering
    \includegraphics[width=\linewidth]{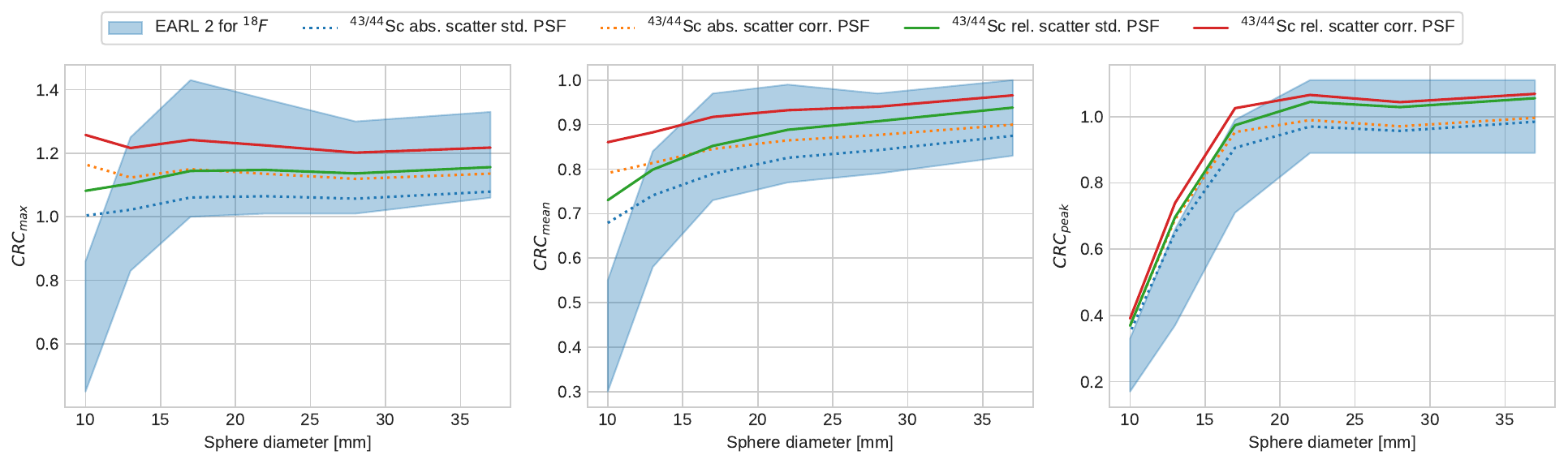} \\
    \vspace{10pt}
    \includegraphics[width=\linewidth]{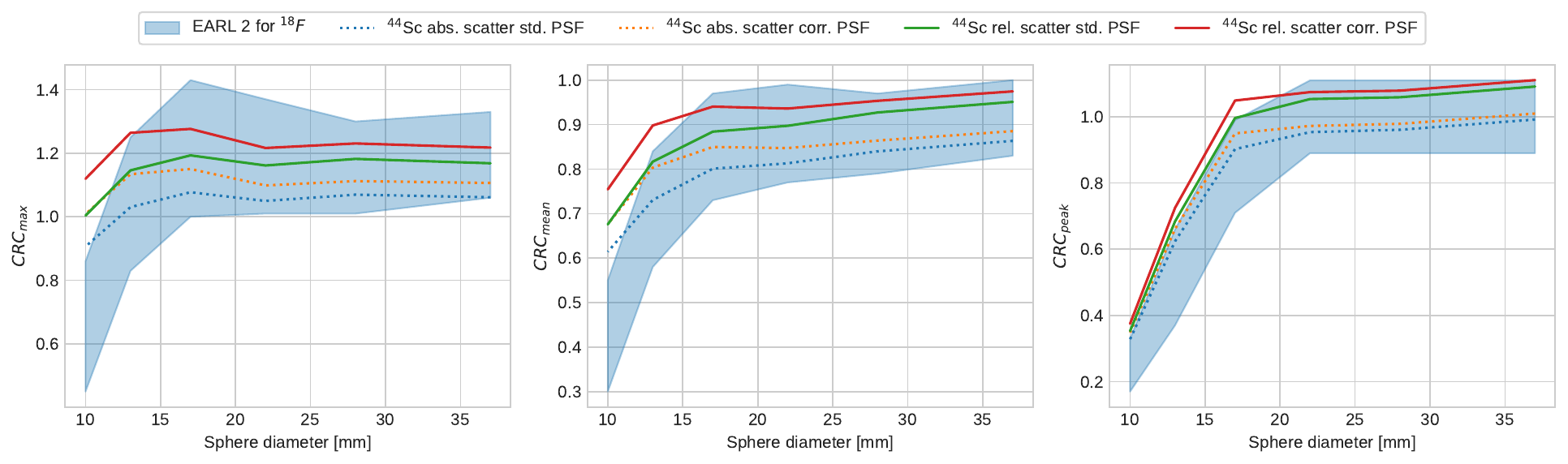}
    \caption{Contrast recovery curves for \scm\ (\emph{top row}) and \scp\ (\emph{bottom row}) for three iterations and no post-reconstruction filter.}
    \label{fig:recovery_curves}
\end{figure}

Fig.~\ref{fig:recovery_curves} shows the three recovery curves $\crmax$, $\crmean$ and $\crpeak$ for \scm\ and \scp\ in comparison with the EARL standard 2 for \isotope[18]{F}. The different recovery curves show relative vs.\ absolute scatter correction as well as the standard vs.\ positron range-corrected PSF. Tabs.~\ref{tab:sc43_recovery} and \ref{tab:sc44_recovery} summarize the corresponding numerical values of the recoveries as well as the COV and lung error.    

\begin{table}[]
    \centering
    \caption{Contrast recovery values, COV and lung error for \scm\ (three iterations and no post-reconstruction filter). }\label{tab:sc43_recovery}
    {\footnotesize
    \begin{tabular}{llcccccccc}
        \toprule
        Recovery & Reconstruction & \multicolumn{6}{c}{Single sphere contrast recovery}  & COV [\%] & Lung err. [\%] \\
        & & $\emptyset10$ & $\emptyset13$ & $\emptyset17$ & $\emptyset22$ & $\emptyset25$ & $\emptyset37$ &  &  \\
        \midrule 
        \rowcolor{lightgray}   & \scm\ abs.\ std.\ PSF  & 1.00 & 1.02 & 1.06 & 1.06 & 1.06 & 1.08 & 11.97 & 9.54 \\ 
        \rowcolor{lightgray}   & \scm\ abs.\ corr.\ PSF & 1.16 & 1.12 & 1.15 & 1.14 & 1.12 & 1.14 & 12.61 & 9.46 \\ 
        \rowcolor{lightgray}   & \scm\ rel.\ std.\ PSF  & 1.08 & 1.10 & 1.14 & 1.15 & 1.14 & 1.16 & 12.23 & 6.05 \\ 
        \rowcolor{lightgray} \multirow{-5}*{$\crmax$}  & \scm\ rel.\ corr.\ PSF & 1.26 & 1.22 & 1.24 & 1.22 & 1.20 & 1.22 & 12.92 & 5.97 \\ 
        & \scm\ abs.\ std.\ PSF  & 0.68 & 0.74 & 0.79 & 0.83 & 0.84 & 0.87 & 11.97 & 9.54 \\ 
        & \scm\ abs.\ corr.\ PSF & 0.79 & 0.81 & 0.85 & 0.86 & 0.88 & 0.90 & 12.61 & 9.46 \\ 
        & \scm\ rel.\ std.\ PSF  & 0.73 & 0.80 & 0.85 & 0.89 & 0.91 & 0.94 & 12.23 & 6.05 \\ 
        \multirow{-5}*{$\crmean$} & \scm\ rel.\ corr.\ PSF & 0.86 & 0.88 & 0.92 & 0.93 & 0.94 & 0.97 & 12.92 & 5.97 \\ 
        \rowcolor{lightgray}   & \scm\ abs.\ std.\ PSF  & 0.35 & 0.65 & 0.91 & 0.97 & 0.96 & 0.98 & 11.97 & 9.54 \\ 
        \rowcolor{lightgray}   & \scm\ abs.\ corr.\ PSF & 0.37 & 0.69 & 0.95 & 0.99 & 0.97 & 1.00 & 12.61 & 9.46 \\ 
        \rowcolor{lightgray}   & \scm\ rel.\ std.\ PSF  & 0.37 & 0.70 & 0.97 & 1.04 & 1.03 & 1.06 & 12.23 & 6.05 \\ 
        \rowcolor{lightgray} \multirow{-5}*{$\crpeak$} & \scm\ rel.\ corr.\ PSF & 0.39 & 0.74 & 1.03 & 1.07 & 1.04 & 1.07 & 12.92 & 5.97 \\ 
        \bottomrule
    \end{tabular}
    }
\end{table}

\begin{table}[]
    \centering
    \caption{Contrast recovery values, COV and lung error for \scp\ (three iterations and no post-reconstruction filter). }\label{tab:sc44_recovery}
    {\footnotesize
    \begin{tabular}{llcccccccc}
        \toprule
        Recovery & Reconstruction & \multicolumn{6}{c}{Single sphere contrast recovery}  & COV [\%] & Lung err. [\%] \\
        & & $\emptyset10$ & $\emptyset13$ & $\emptyset17$ & $\emptyset22$ & $\emptyset25$ & $\emptyset37$ &  &  \\
        \midrule 
        \rowcolor{lightgray}   & \scp\ abs.\ std.\ PSF  & 0.91 & 1.03 & 1.08 & 1.05 & 1.07 & 1.06 & 10.45 & 11.88 \\ 
        \rowcolor{lightgray}   & \scp\ abs.\ corr.\ PSF & 1.01 & 1.13 & 1.15 & 1.10 & 1.11 & 1.11 & 11.09 & 11.85 \\ 
        \rowcolor{lightgray}   & \scp\ rel.\ std.\ PSF  & 1.00 & 1.15 & 1.19 & 1.16 & 1.18 & 1.17 & 10.78 & 5.57 \\ 
        \rowcolor{lightgray} \multirow{-5}*{$\crmax$}  & \scp\ rel.\ corr.\ PSF & 1.12 & 1.26 & 1.28 & 1.22 & 1.23 & 1.22 & 11.48 & 5.53 \\ 
        & \scp\ abs.\ std.\ PSF  & 0.61 & 0.73 & 0.80 & 0.81 & 0.84 & 0.86 & 10.45 & 11.88 \\ 
        & \scp\ abs.\ corr.\ PSF & 0.68 & 0.80 & 0.85 & 0.85 & 0.86 & 0.89 & 11.09 & 11.85 \\ 
        & \scp\ rel.\ std.\ PSF  & 0.68 & 0.82 & 0.88 & 0.90 & 0.93 & 0.95 & 10.78 & 5.57 \\ 
        \multirow{-5}*{$\crmean$} & \scp\ rel.\ corr.\ PSF & 0.75 & 0.90 & 0.94 & 0.94 & 0.95 & 0.97 & 11.48 & 5.53 \\ 
        \rowcolor{lightgray}   & \scp\ abs.\ std.\ PSF  & 0.33 & 0.62 & 0.90 & 0.95 & 0.96 & 0.99 & 10.45 & 11.88 \\ 
        \rowcolor{lightgray}   & \scp\ abs.\ corr.\ PSF & 0.35 & 0.66 & 0.95 & 0.97 & 0.98 & 1.01 & 11.09 & 11.85 \\ 
        \rowcolor{lightgray}   & \scp\ rel.\ std.\ PSF  & 0.35 & 0.68 & 1.00 & 1.05 & 1.06 & 1.09 & 10.78 & 5.57 \\ 
        \rowcolor{lightgray} \multirow{-5}*{$\crpeak$} & \scp\ rel.\ corr.\ PSF & 0.37 & 0.73 & 1.05 & 1.07 & 1.08 & 1.11 & 11.48 & 5.53 \\ 
        \bottomrule
    \end{tabular}
    }
\end{table}

\begin{figure}
    \centering
    \includegraphics[width=\linewidth]{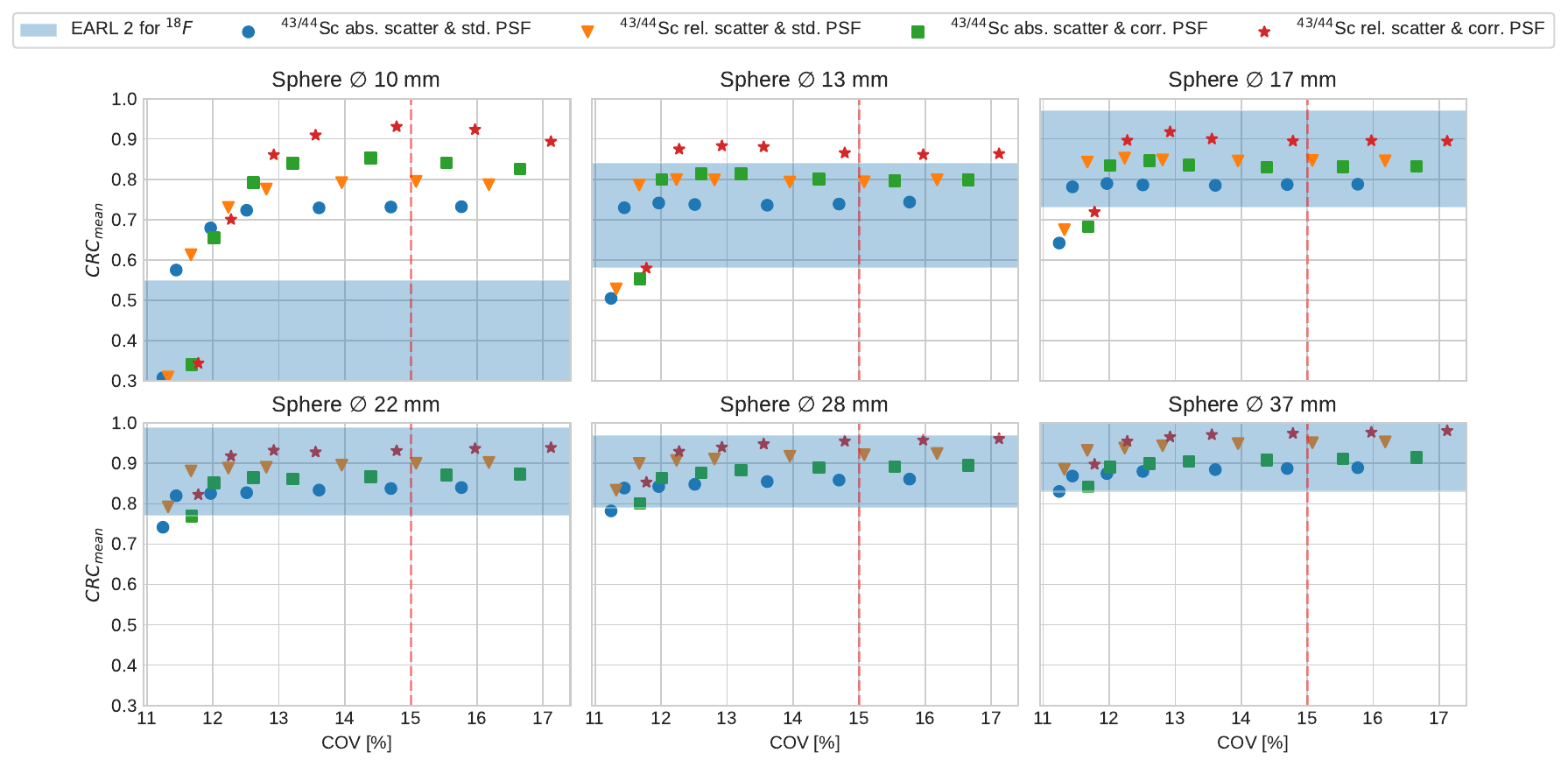}
    \caption{$\crmean$ as a function of COV for \scm\ for an increasing number of iterations and different scatter corrections. EARL limits are shown as the dashed line and the colored band. }
    \label{fig:sc43_convergence}
\end{figure}

\begin{figure}
    \centering
    \includegraphics[width=\linewidth]{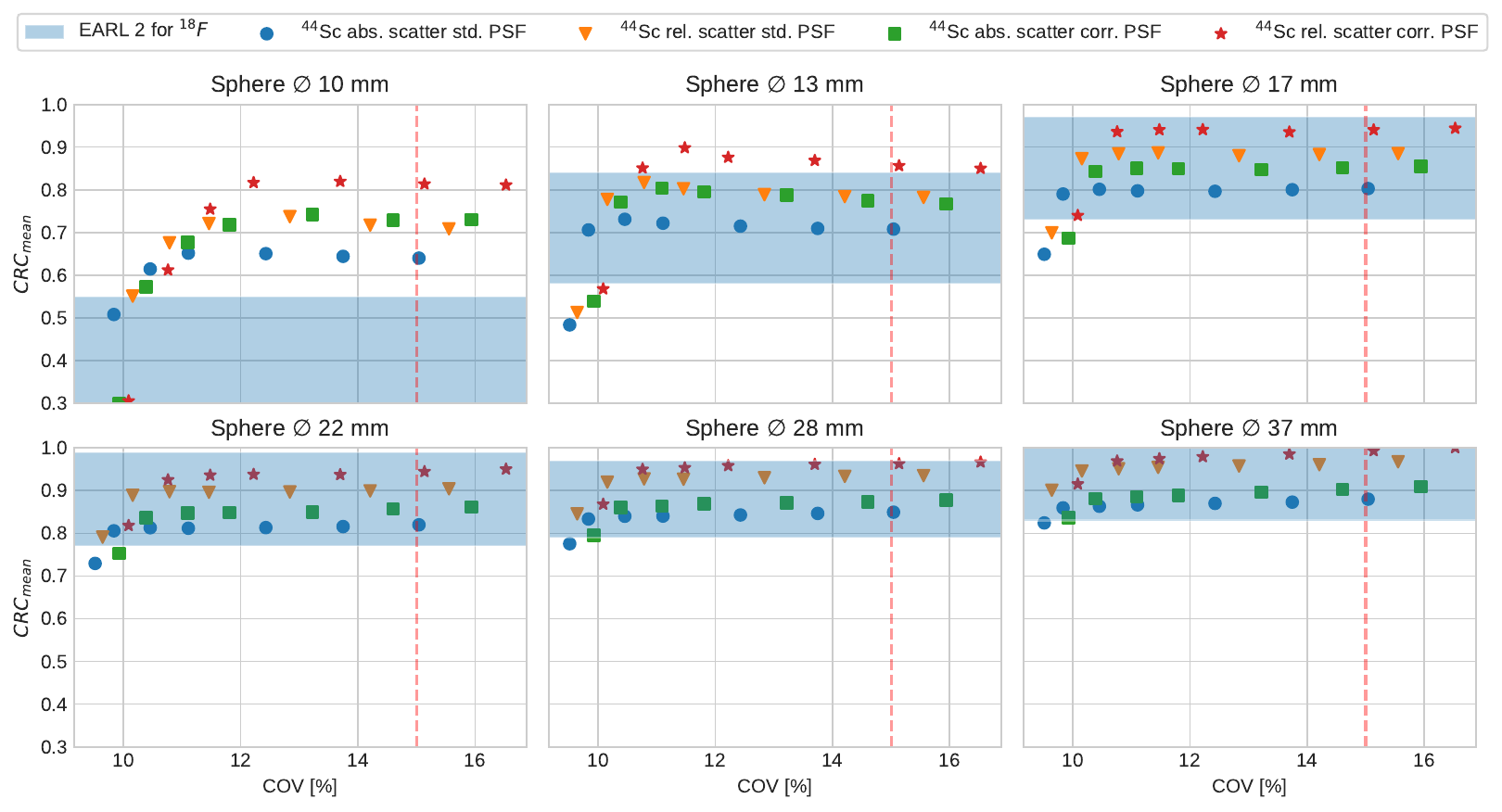}
    \caption{$\crmean$ as a function of COV for \scp\ for an increasing number of iterations and different scatter corrections. EARL limits are shown as the dashed line and the colored band.}
    \label{fig:sc44_convergence}
\end{figure}

\begin{figure}
    \centering
    \includegraphics[width=0.85\linewidth]{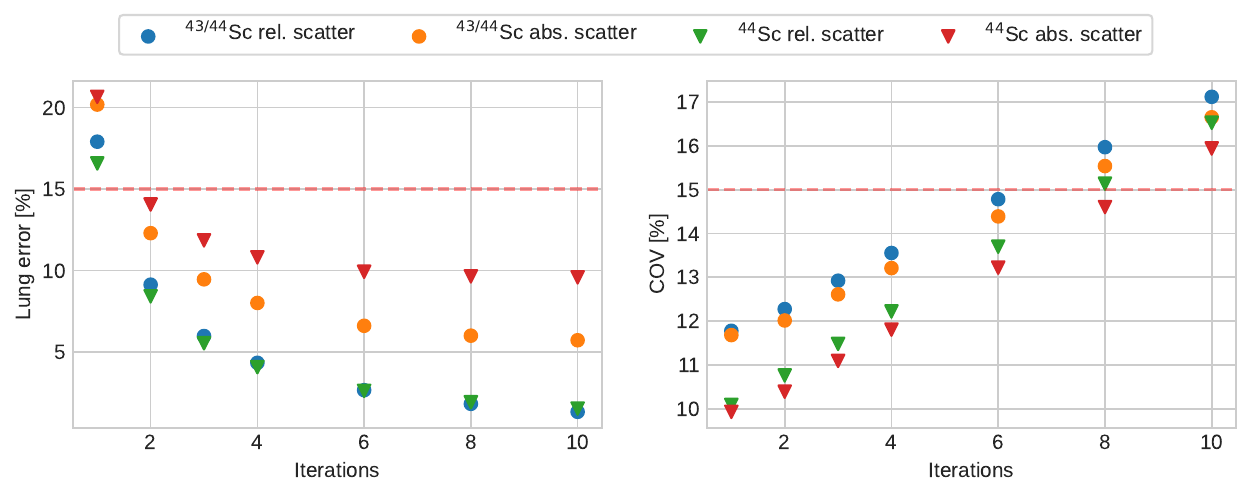}
    \caption{Lung error and COV as a function of the number of iterations with the corrected PSF and no post-reconstruction filter. The red dashed line indicates the EARL limits.}
    \label{fig:cov_lung}
\end{figure}

In Figs.~\ref{fig:sc43_convergence} and \ref{fig:sc44_convergence} we present the convergence of the reconstruction algorithm with increasing number of iterations. The $\cov$ is plotted against $\crmean$ for each sphere of the NEMA phantom. The accepted limit of $15\%$ for the COV is marked with the dashed line while the colored band depicts the EARL standard 2 limits.
Related to the convergence of the OSEM reconstruction algorithm, Fig.~\ref{fig:cov_lung} shows the lung error and the COV as a function of the number of iterations. As for the recovery curves, the number of subsets is fixed to five and there is no post-reconstruction filtering. 

\begin{figure}
    \centering
    \includegraphics[width=0.85\linewidth]{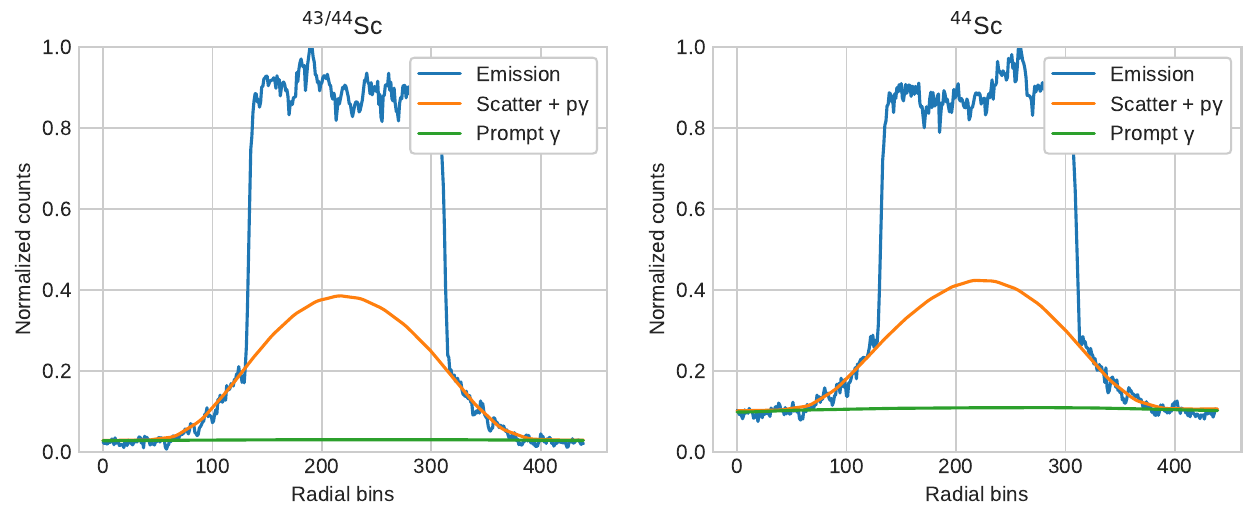}
    \caption{Trues sinogram together with total correction and \pg\ correction sinograms at $0^\circ$ view angle.}
    \label{fig:sinograms}
\end{figure}

Finally, Fig.~\ref{fig:sinograms} shows the normalized trues sinograms for \scm\ and \scp\ at $0^\circ$ view angle in the case of relative scatter correction. The total correction sinogram, i.e.\ scatter and the \pg\ contribution, is plotted alongside the \pg\ correction illustrating the difference in (relative) \pg\ contribution in the tail regions of the two sinograms. 

\section{Discussion}

Our study confirms the expected quantitative imaging performance of both \scm\ and \scp\ to be comparable to \isotope[18]{F}. Although this was already shown in Refs.~\cite{lima2020,lima2021} for a clinical PET/CT scanner, we established that the scandium radioisotopes meet EARL's updated contrast recovery limits when imaged with a LAFOV PET / CT. Furthermore, we showed how a radioisotope-specific PSF correction can significantly improve the contrast recovery. 

Our main concern is the rather large bias, in particular for \scm. The $10\%$ limit of Ref.~\cite{Sunderland2025} for radionuclides other than \isotope[18]{F} and \isotope[68]{Ga} is met for \scp. Likely, the $B_{\scp}=9.6\%$ bias we obtained for \scp\ could be improved by implementing a calibration factor from the vendor or from the national metrology institute in Insel's dose calibrator. The bias $B_{\scm} =-17.1\%$, however, lies outside the allowed range. The total activity in the aliquot measured by the HPGe spectrometer (including \isotope[44m]{Sc}) exceeds the activity measured with the dose calibrator (both measurements at PSI) by $18.5\%$. This might be seen already as an indication that a mixture-specific calibration factor for the dose calibrator is necessary. Although the \isotope[44m]{Sc} activity is small compared to \isotope[43]{Sc} and \scp\ (grows to $2\%$ at scan time due to the long half-life), the additional prompt photons might offset the dose calibrator measurement. Furthermore, determining the scaling factor between Insel's and PSI's dose calibrators based on a simple cross-calibration bears a rather big risk of increasing the bias. 
Ref.~\cite{lima2021} does not specify the details of the definition of \SI{}{\becquerel} for the \scm\ mixture. They cross-calibrate the local dose calibrator with the reference activity from PSI's dose calibrator. No details of positron fraction corrections or treatment of the exact mixture are given. For a clinical translation of \scm\ a correct gauging of a dose calibrator by a metrology institute for the produced isotope mixture and the implementation of the mixture parameters in the scanner software would be necessary. Until then, we believe that an absolute image quantification will remain challenging for \scm. Beyond quantification, the limited radionuclidic purity of \scm\ poses a challenge for its clinical translation, since the isotopic composition varies over time and depends on details of the production process).


The physical decay properties of \isotope[43]{Sc} would suggest superior imaging properties compared to \scp. The positron energy is smaller (see Fig.~\ref{fig:positron_range}) and the \pg\ emission is much smaller and of lower energy (main line at \SI{372.9\pm0.3}{\kilo\electronvolt} at $22.5\pm0.7\%$ intensity vs.\ \SI{1157.022\pm0.015}{\kilo\electronvolt} at $99.887\pm0.003\%$ for \scp). However, we observed that there is only a marginal difference between \scm\ and \scp\ in the recovery curves of Fig.~\ref{fig:recovery_curves}. The main (visible) difference appears in the tail regions of the sinograms in Fig.~\ref{fig:sinograms}. The tails are caused mainly by \pg\ false coincidences \cite{Conti2016}. The ratio of the (almost constant) sinogram tails relative to the maximum between \scm\ and \scp\ was around $27\%$. This is consistent with the $27.8\%$ \scp\ in the \scm\ mixture, indicating that the \pg\ randoms are caused by the \scp\ in the mixture. The branching ratio for \pg\ within or above the coincidence energy window is below $0.02\%$ for \isotope[43]{Sc} and will therefore not really contribute to the random coincidences outside the source regions.  

As shown also in Ref.~\cite{Mercolli2026b}, the vendor's \pg\ correction algorithm is effective on Quadra. This is clearly visible when considering the absolute vs.\ relative scatter correction. The absence of a \pg\ correction brings $\crmax$ closer to $1.0$ compared to the case of relative scatter correction, but clearly deteriorates $\crmean$. Most importantly, however, the missing \pg\ correction in absolute scatter correction leads to a larger lung error (see Fig.~\ref{fig:cov_lung}). It should also be noted that the lung error is significantly larger for \scp\ in the absence of a \pg\ correction. For relative scatter, the lung error is hardly distinguishable for \scp\ and \scm, indicating an effective correction of \pg\ randoms. Absolute scatter correction without \pg\ correction should therefore be avoided in any clinical protocol for scandium isotopes on Quadra. 


Figs.~\ref{fig:sc43_convergence} and \ref{fig:sc44_convergence} show a very fast convergence of the OSEM reconstruction algorithm. Mostly independent of the isotope, scatter or PSF correction, there is little improvement in $\crmean$ after three iterations, even for the $\emptyset \SI{10}{\mm}$ phantom sphere. Note that clinical \isotope[18]{F} protocols on Quadra typically use four iterations \cite{Prenosil2022a}. Of course, four iterations further decrease the lung error (see Fig.~\ref{fig:cov_lung}). Depending on whether one is willing to apply a post-reconstruction filter (improving COV), the loss in COV may outweigh the gain in lung error at mostly equal contrast recovery. We therefore advocate that three iterations should be used for a clinical protocol with scandium isotopes in the case of an all-pass filtering. We refrain from exploring the reconstruction parameter space with different numbers of subsets but rely on the clinical standard of five subsets \cite{Prenosil2022a}. 

While Refs.~\cite{lima2020,lima2021} relied on the standard PSF correction in the vendor's reconstruction algorithm, we apply an isotope-specific PSF correction. In Tabs.~\ref{tab:sc43_recovery} and \ref{tab:sc44_recovery} we note that $\crmean$ improves up to $0.13$ for \scm\ and $0.08$ for \scp\ in the case of relative scatter correction. A similar improvement in contrast recovery was observed by Ref.~\cite{Kertesz2022} in the case of \isotope[124]{I}, which has a comparable positron range in water to that of \scp. Note, however, that we do not implement isotope-specific projectors in the OSEM algorithm as done in Ref.~\cite{Kertesz2022}. This is also the reason why we do not observe a significant difference in the convergence (see Figs.~\ref{fig:sc43_convergence} and \ref{fig:sc44_convergence}). Since we apply a post-reconstruction deconvolution, $\crmax$ is expected to increase as well as the COV. Since the ROI position for $\crpeak$ is optimized while keeping the size constant, the PSF correction has only a marginal impact. Though the clinical standard is still without nuclide-specific PSF modeling (even for \isotope[82]{Rb} \cite{Lassen2025}), we hope that vendors will implement (material-dependent) positron range corrections for standard as well as novel radionuclides.

In summary, we confirmed the findings of Refs.~\cite{lima2020,lima2021} that clinical PET/CT systems can image scandium isotopes with an image quality that meets the EARL standard 2 criteria for \isotope[18]{F}. The vendor's corrections applied in the image reconstruction process are effective for both \scm\ mixture and pure \scp, but could be improved with an isotope-specific PSF correction. The increased sensitivity of LAFOV scanners in conjunction with the longer half-life of \isotope[43]{Sc} and \scp\ could allow for longer uptake times \cite{Alberts2021b}. Despite the challenges in producing \scm\ with a medical cyclotron \cite{Carzaniga2017}, challenging radionuclidic composition and the difficulties with the bias, we would like to motivate further investment in research and development towards the clinical translation of \scm, given its favorable imaging properties and significantly reduced radiation dose to a patient compared to \scp\ \cite{Gomes2024}.


\section{Conclusion}


In this study, we advanced the quantitative imaging of \scm\ and \scp\ on a clinical LAFOV PET/CT scanner. Using a NEMA phantom, we confirmed that the imaging performance of both scandium isotopes is comparable to \isotope[18]{F} and meets the updated EARL standard 2 contrast recovery criteria. We showed that the vendor's \pg\ correction, is essential to control the random coincidences and lung error introduced by the \pg\ emission of scandium isotopes, and that an isotope-specific PSF correction further improves contrast recovery. Based on the convergence behavior of the reconstruction, we recommend three iterations with relative scatter correction (including \pg\ correction) as a practical clinical protocol on Quadra. The main outstanding challenge is the bias observed for the \scm\ mixture. Together with the longer half-lives of \isotope[43]{Sc} and \scp\ compared to \isotope[18]{F} and \isotope[68]{Ga}, the increased sensitivity of LAFOV systems opens the possibility of longer uptake times and/or multi-time-point imaging, supporting the clinical translation of scandium-based theranostic pairs.

\section*{Acknowledgments}

The authors thank Hasan Sari, William M. Steinberger for useful discussions and for their help with e7-tools. They also acknowledge the help of Ângelo Rafael Felgosa Cardoso, Angela Filipa Silva Mendes and Marco Viscione for the scheduling of the experiments at Inselspital.

\section*{Declarations}

\subsection*{Funding}
\noindent This research is partially supported by the grant no. 216944 of the Swiss National Science Foundation. The dangerous good transportation was financed by the Research Fund of the Swiss Society of Radiobiology and Medical Physics.

\subsection*{Competing interests}
\noindent MT is a full-time employee of Siemens Medical Solutions USA Inc. All other authors have no conflicts of interests to report.

\subsection*{Ethics approval}
\noindent Not applicable. 

\subsection*{Consent to participate}
\noindent Not applicable. 

\subsection*{Data availability}
\noindent The datasets generated and/or analyzed in the current study are available from the corresponding author upon reasonable request.

\subsection*{Authors' contributions}
\textbf{LM}: Conceptualization, Formal analysis, Investigation, Methodology, Software, Validation, Writing - Original Draft, Writing - Review \& Editing
\textbf{MT}: Software, Methodology, Writing - Review \& Editing
\textbf{AEG}: Software, Formal analysis, Writing - Review \& Editing
\textbf{PVG and ANM:} Investigation, Resources, Writing - Review \& Editing
\textbf{SB:} Conceptualization, Project administration, Writing - Review \& Editing
\textbf{NPvdM:} Conceptualization, Investigation, Supervision, Project administration, Writing - Review \& Editing

\printbibliography

\end{document}